\documentclass[showpacs,preprintnumbers,amsmath,amssymb,twocolumn]{revtex4} 
 
\usepackage{graphicx}
\usepackage{dcolumn}
\usepackage{bm}

\def\openone{\leavevmode\hbox{\small1\kern-3.8pt\normalsize1}}%
 
\def\mf{{\mbox{\tiny\em MFA}}}

\def\bea{\begin{eqnarray}} 
\def\eea{\end{eqnarray}} 
\def\beq{\begin{equation}} 
\def\eeq{\end{equation}}

\begin{document} 
 
\title{Color Neutral Ground State of 2SC Quark Matter} 
\author{D. Blaschke$^{a,b}$, D. G\'omez Dumm$^{c,d}$, 
A. G. Grunfeld$^{e}$ and N. N. Scoccola$^{d,e,f}$} 
\affiliation{$^a$ Gesellschaft f\"ur Schwerionenforschung (GSI) mbH, 
Planckstr. 1, D-64291 Darmstadt, Germany\\ 
$^b$ Bogoliubov Laboratory for Theoretical Physics, JINR Dubna, 141980 Dubna, 
Russia\\ 
$^c$ IFLP,CONICET - Depto. de F\'isica, FCE, Universidad Nacional de La Plata, 
C.C. 67, 1900 La Plata, Argentina\\ 
$^d$ CONICET, Rivadavia 1917, 1033 Buenos Aires, Argentina\\ 
$^e$ Physics Department, Comisi\'on Nacional de Energ\'ia At\'omica, 
Av. Libertador 8250, 1429 Buenos Aires, Argentina\\ 
$^f$ Universidad Favaloro, Sol\'is 453, 1078 Buenos Aires, Argentina}

\begin{abstract} 
We construct a new color neutral ground state of two-flavor color 
superconducting quark matter. It is shown that, in contrast with the 
conventionally considered ground state with diquark pairing in only one 
color direction, this new state is stable against arbitrary diquark 
fluctuations. In addition, the thermodynamical potential is found to 
be lower for this new state than for the conventional one. 
\end{abstract} 
 
\date{\today} 
 
 
\maketitle 
 
Recent investigations on the QCD phase diagram have discovered a rich 
diversity of color superconducting quark matter phases at low temperatures 
and intermediate densities \cite{cscreview}. 
Particularly interesting are possible implications of 
these results for the physics of compact stars \cite{cscstars} as well as 
heavy ion collision experiments \cite{cschic} addressing the domain of 
densities and temperatures where the strange quarks are still heavy and 
confined \cite{Ruster:2005jc,Blaschke:2005uj} . In both 
systems, the constraint of global color neutrality has to be fulfilled. In 
addition, the constraint of global electric neutrality has to be satisfied 
when macroscopic objects like compact stars are considered and flavor 
changing processes have enough time to adjust $u$ and $d$ quark chemical 
potentials according to $\beta$-equilibrium. 
 
 
In view of the nonperturbative character of QCD, the theoretical 
treatment of hadronic matter at the vicinity of the phase 
transitions for low temperatures and finite densities is a problem 
of highest complexity, where rigorous theoretical approaches are 
not yet available and lattice QCD simulations are up to now not 
applicable. Therefore one has to rely on effective 
field-theoretical models of interacting quark matter, which are 
built taking into account the symmetry requirements of the QCD 
lagrangian and offer the possibility of dealing with the yet 
simplified interactions in a systematic way. Chiral quark models 
of QCD that adopt a current-current form of the interaction with 
mesonic and diquark components have been particularly useful, 
since the theories can be bosonized in a straightforward way. In 
the meson-diquark representation of these models an effective 
quantum hadrodynamics can be derived \cite{cahill}, but it has not 
yet been used to explore the QCD phase diagram. A preparatory step 
for this formidable task is the investigation of the mean field 
approximation (MFA), where important progress has been recently 
made within a nonlocal, covariant formuation that has been 
extended even to the study of color superconductivity in the QCD 
phase diagram~\cite{Blaschke:2004cc}. 
 
 
The two-flavor color superconductivity (2SC) phase of quark matter 
has been first considered in instanton-motivated QCD models in 
\cite{Rapp:1997zu}, where the question of color neutrality was 
disregarded. However, it was soon realized that the 2SC phase in 
which color symmetry is broken by the orientation of the diquark 
field in one of the color directions (2SC-$b$) entails a mismatch 
in the quark densities of paired and unpaired colors provided that 
color chemical potentials are all equal to each other, 
$\mu_r=\mu_g=\mu_b$. Therefore, in the 2SC-$b$ state color 
neutrality requires color charge chemical potentials to be 
readjusted so that $\mu_8=\frac12 (\mu_r+\mu_g-2\mu_b)$ acquires a 
nonvanishing value while $\mu_3=\frac32 (\mu_r-\mu_g)$ remains 
zero due to the degeneracy of the red and green colors. While this 
adjustment of $\mu_8\neq 0$ has long been considered a proper 
solution of the color neutrality constraint \cite{Huang:2002zd}, a 
recent investigation of fluctuations around the mean field 
oriented in the blue (0,0,1) direction has revealed the 
instability of this state once color neutrality is required 
\cite{hjz}. In the present paper we investigate the entire space 
of mean-field orientations in order to look for color neutral 
states which are stable against fluctuations. We find that these 
correspond to color neutral symmetric states (2SC-s) for which the 
condensates are equal in modulus in all the three directions of the 
color space. 
 
As in Ref.~\cite{hjz}, we consider the simplest version of the 
flavor SU(2) Nambu$-$Jona-Lasinio model \cite{njl,njlquark,klev}, 
extended so as to include the quark-quark interaction sector and 
finite chemical potentials 
\begin{eqnarray} 
\label{njl} {\cal L} &=& 
\bar{\psi}\left(i\gamma^{\mu}\partial_{\mu}+\mu\gamma_0 - \hat 
m\right)\psi 
+G_S\!\left[\!\left(\bar{\psi}\psi\right)^2\!+\left(\bar{\psi}i\gamma_5\vec{ 
\tau}\psi\right)^2\right]\nonumber\\ 
&+&\! G_D\left(\bar\psi^c_{i\alpha} 
i\gamma^5\epsilon^{ij}\epsilon^{\alpha\beta\gamma}\psi_{j\beta}\right) 
\left(\bar\psi_{i\alpha} 
i\gamma^5\epsilon^{ij}\epsilon^{\alpha\beta\gamma}\psi^c_{j\beta}\right) 
\ . 
\end{eqnarray} 
Here $\hat m$ is the diagonal current mass matrix for light 
quarks, $G_S$ and $G_D$ are coupling constants in color singlet 
and anti-triplet channels respectively, and  $\psi_{i\alpha}$ 
stands for quark fields with flavor index $i=u,d$ and color index 
$\alpha=r,g,b$ (charge conjugated fields are given by 
$\psi_{i\alpha}^c = i\gamma^2\gamma^0\bar\psi_{i\alpha}^T$). The 
Pauli matrices ${\vec\tau} =(\tau_1, \tau_2, \tau_3)$ act in 
flavor space, while and $\epsilon^{ij}$ and 
$\epsilon^{\alpha\beta\gamma}$ are totally antisymmetric tensors 
in flavor and color spaces, respectively. In the present letter we 
are mainly concerned with the effect of color neutrality on the 
ground state of a two-flavor color superconductor, therefore we 
will restrict here the discussion to the flavor symmetric case and 
consider the extension to electrically neutral matter elsewhere. 
Regarding quark chemical potentials, the elements of the matrix 
\begin{equation} 
\label{mu1} 
\mu={\rm diag}\,(\mu_{r},\mu_{g},\mu_{b},\mu_{r},\mu_{g},\mu_{b}) 
\end{equation} 
can be written as 
\begin{eqnarray} 
\label{mu2} 
&&\mu_{r}=\mu_B/3+\mu_8/3+\mu_3/3\ ,\nonumber\\ 
&&\mu_{g}=\mu_B/3+\mu_8/3-\mu_3/3\ ,\nonumber\\ 
&&\mu_{b}=\mu_B/3-2\mu_8/3\ , 
\end{eqnarray} 
where $\mu_B$ is the baryon chemical potential, while $\mu_8$ and 
$\mu_3$ are introduced to ensure color charge neutrality. Our aim 
is to discuss the color superconducting phase of the model in the 
mean-field approximation, which in general is characterized by 
nonvanishing diquark condensates 
\begin{equation} 
\label{delta} 
\Delta_\alpha = -\,2\,G_D\,\langle 
\bar\psi^c_{i\beta}i\gamma^5\epsilon^{ij} 
\epsilon^{\beta\gamma\alpha}\psi_{j\gamma}\rangle, 
~~~\alpha=r, g, b\ . 
\end{equation} 
For standard values of the diquark coupling, $G_D=\frac34 G_S$, 
there is no simultaneous chiral symmetry breaking in this phase. 
In addition, since for light quarks the current quark masses are 
significantly smaller than the typical values of $\mu_B$ and 
$\Delta_\alpha$, for the purpose of the present study we can 
safely neglect both these small masses and the corresponding 
mesonic mean fields. Within this limit, we proceed to calculate 
the thermodynamical potential per unit volume at zero temperature. 
For convenience we perform our calculations in Euclidean space, 
where the thermodynamical potential in MFA is given by 
\begin{equation} 
\Omega^\mf \ = \ \frac{\Delta^2}{4 G_D} - \frac{1}{2} \int 
\frac{d^4 p}{(2\pi)^4} \ \ln \; {\rm det}\; M \ , 
\label{OmegaMFA} 
\end{equation} 
and the constraint of global color neutrality has to be obeyed, 
i.e. color charge densities should vanish 
\begin{equation}
\label{Qvan}
Q_\alpha=-\partial \Omega / \partial \mu_\alpha = 0~, ~~\alpha=3,8~.
\end{equation}
Here we have defined $\Delta^2 = \sum_{\alpha=r,g,b} 
\Delta_\alpha^2$, and the space integral is regulated as usual by 
introducing a sharp three-momentum cutoff $\Lambda$. The inverse 
fermion propagator $M$ is a $48\times 48$ matrix in Nambu-Gorkov, 
Dirac, color and flavor spaces, and can be conveniently written as 
\begin{equation} 
\label{M} 
M=\left(\begin{array}{cc} 
M^+& 0\\ 
0& M^- 
\end{array}\right)\ , 
\qquad M^- = - {M^+}^\dagger \ , 
\end{equation} 
with 
\begin{widetext} 
\begin{equation} 
\label{S+} 
M^+ \ = \ \left(\begin{array}{cccccc} 
(G^+_{0r})^{-1}& 0& 0& 0& -\Delta_b'& \Delta_g'\\ 
0& (G^+_{0g})^{-1}& 0& \Delta_b'& 0& -\Delta_r'\\ 
0& 0& (G^+_{0b})^{-1}& -\Delta_g'& \Delta_r'& 0\\ 
0& \Delta_b'& -\Delta_g'& (G^-_{0r})^{-1}& 0& 0\\ 
-\Delta_b'& 0& \Delta_r'& 0& (G^-_{0g})^{-1}& 0\\ 
\Delta_g'& -\Delta_r'& 0& 0& 0& (G^-_{0b})^{-1} 
\end{array}\right)~, 
\end{equation} 
\end{widetext} 
where $\Delta_{\alpha}'=i\gamma_5 \Delta_{\alpha}$ and 
$(G^\pm_{0\alpha})^{-1}= -[(p_4 \mp i\mu_{\alpha})\gamma_4 
+\vec{p}\cdot \vec{\gamma}]$ are $4\times 4$ matrices in Dirac 
space. After some algebra, it is seen that the determinant can be 
cast into the form 
\begin{equation} 
{\rm det}\; M = (\; S^+\; S^-\,)^4\ , 
\end{equation} 
where 
\begin{eqnarray} 
S^{\pm} &=& |C_r^{\pm} C_g^{\pm} C_b^{\pm}|^2 
+ |C_r^{\pm} \Delta_r^2 + C_g^{\pm} \Delta_g^2 + C_b^{\pm} \Delta_b^2|^2 
 \nonumber\\ 
&& \hspace{-.8cm} 
+ \ 2 \!\left[ |C_r^{\pm}|^2 \, {\rm Re}\,({C_g^{\pm}}^\ast 
C_b^{\pm}) \; \Delta_r^2 + {\rm cycl. perm. \{{\it rgb}\}}\right] 
\! , 
\label{ese} 
\end{eqnarray} 
with 
\[ 
C_{\alpha}^\pm = \mu_{\alpha}\; \pm \; |\vec p\,| \; + \; i\,p_4 \ . 
\] 
With this general expression at hand we can investigate the 
problem of finding the most favored state under the constraint of 
the color neutrality, i.e. $Q_3=Q_8=0$.
 
Since the expression in Eq.(\ref{ese}) is totally symmetric under cyclic 
permutations of the three color indices, a cubic symmetry is expected in 
the three-dimensional space spanned by the three color directions along 
which the magnitudes of the colored diquark condensates are the 
coordinates. In the standard Cartesian representation, the condensate 
vector in color space is given by 
\begin{equation} 
\vec{\Delta}=\Delta_r~ \vec{\rm e}_r 
+\Delta_g~ \vec{\rm e}_g +\Delta_b~ \vec{\rm e}_b~. 
\end{equation} 
Due to the above mentioned symmetry, in what follows we will 
restrict our discussion to the sector defined by nonnegative 
values of the Cartesian coordinates $\Delta_\alpha$. In this 
representation, the 2SC-$b$ state is given by $\vec{\Delta}_{{\rm 
2SC-}b}=(0, 0, \Delta_b)$, whereas the color symmetric state is 
$\vec{\Delta}_{\rm 2SC-s}=(\Delta_s, \Delta_s, \Delta_s)$, i.e.\ a 
vector pointing from the center to one of the edges of a cube in 
color space. For the 2SC-s state color neutrality is achieved with 
$\mu_8=\mu_3=0$, while everywhere else color symmetry is broken, 
entailing that either $\mu_8$, $\mu_3$ or both have to be 
different from zero in order to fulfill the color neutrality 
constraint. 
 
The observation made in \cite{hjz} stating that a 2SC-$b$ state 
defines a saddle point of the thermodynamical potential 
(\ref{OmegaMFA}) in the order parameter space, being a minimum in 
the blue direction but a maximum in the red and green ones, leads 
to the important fact that the 2SC-$b$ state widely considered in 
the literature is not the true ground state of quark matter in the 
2SC phase. Now the problem is to find the true 2SC ground state, 
which should be thermodynamically more favorable not only by a 
lower energy but also by its stability with respect to 
fluctuations in the amplitude and the orientation of the 
condensate. 
\begin{figure}[htb] 
\includegraphics[width=0.4\textwidth]{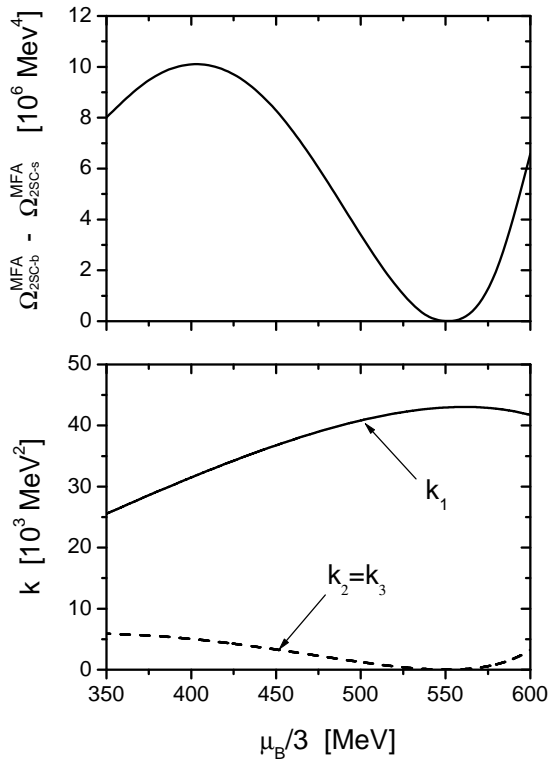} 
\caption{Upper panel: Difference between the thermodynamical 
potentials of two color neutral states, namely the standard one 
--with broken color symmetry-- and the new color symmetric one, as 
functions of the baryon chemical potential $\mu_B$ at temperature 
$T=0$. Lower panel: Eigenvalues $k_1$, $k_2=k_3$ of the curvature 
tensor obtained from the thermodynamical potential at the 
color-symmetric minimum. Strict positivity proves the stability 
w.r.t.\ arbitrary small amplitude fluctuations. Color neutrality 
has been imposed, but not electric charge neutrality.} 
\label{fig1} 
\end{figure} 
We show here that the 2SC-s state proposed in this paper fulfills 
these conditions. To do this, we perform a numerical analysis 
taking a phenomenologically acceptable set of model parameters, 
namely $\Lambda = 653$ MeV, $G_S\Lambda^2=2.14$~\cite{klev}, 
together with $G_D=\frac34 G_S$. First, we show that the 
difference between the thermodynamical potentials of the 2SC-$b$ and 
2SC-s states is positive along the relevant range of baryochemical 
potentials, say $350\ {\rm MeV}\le \mu_B/3\le 600\ {\rm MeV}$. The 
corresponding curve is plotted in the upper panel of Fig.\ 1. 
Second, in the lower panel of Fig.\ 1 we prove the stability of 
the 2SC-s solution by showing the strict positivity of the 
eigenvalues $k_1$, $k_2$, $k_3$ of the curvature tensor 
\begin{equation} 
K_{ij}=\frac{1}{2} \frac{\partial \Omega^\mf}{\partial 
\Delta_i \partial \Delta_j}\bigg|_{\rm 2SC-s} 
\end{equation} 
derived from the thermodynamical potential in the 2SC-s state. 
Interestingly, the largest eigenvalue $k_1$ corresponds to 
fluctuations in the ``radial'' direction $(1,1,1)$, while the two 
eigenvalues denoting the curvature in the orthogonal ``angular'' 
directions are found to be degenerate ($k_2=k_3$). 
 
Note that a special situation occurs at a chemical potential 
$\mu_B/3\simeq 550$ MeV, where the 2SC-s state becomes 
energetically degenerate with the 2SC-$b$ state and simultaneously 
the curvature in the angular directions vanishes. This implies 
that all orientations of the condensate vector $\vec \Delta$ are 
equivalent to each other, i.e.\ the cubic symmetry degenerates to 
a spherical one and the preference of the 2SC-s state gets lost 
(one has in this case $\Delta_b = \sqrt3\Delta_s$). In contrast, 
for all other values of $\mu_B$ the 2SC-s state is preferred, 
owing to the penalty introduced for all other states which need 
finite color chemical potentials to achieve color neutrality. 
\begin{figure}[htb] 
\includegraphics[width=0.45\textwidth]{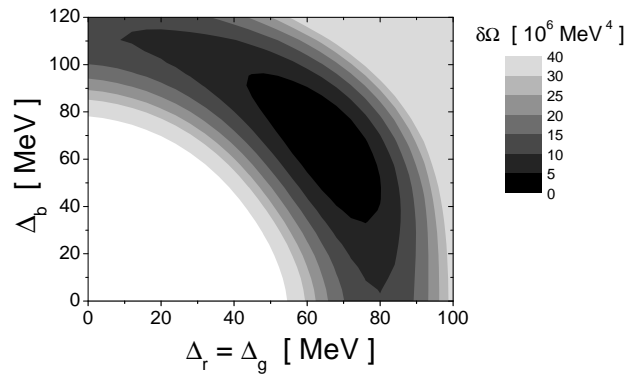} 
\caption{The penalty for the thermodynamical potential, $\delta 
\Omega = \Omega^{MFA} - \Omega^{MFA}_{{\rm 2SC-}s}$, induced by 
the color neutrality constraint in the plane of the order 
parameters $\Delta_b$ and $\Delta_r=\Delta_g$ for $\mu_B=1200$ 
MeV. Since $\mu_3=0$, the penalty is a function of $\mu_8$ and 
both vanish in the 2SC-s state where 
$\Delta_r=\Delta_g=\Delta_b$.} \label{fig2} 
\end{figure} 

The behavior of this penalty $\delta \Omega$ in the condensate state 
space is illustrated by the contour plot shown in Fig.\ 2 for a
baryochemical potential $\mu_B=1200$~MeV chosen such that a maximal
effect can be demonstrated, see Fig. 1. In this plot 
we consider for simplicity the plane spanned by the axes 
$\Delta_r=\Delta_g$ and $\Delta_b$, so that red and green colors 
are degenerate and one has $\mu_3=0$. The penalty, arising from 
color symmetry breaking, is a function of $\mu_8$ and vanish in 
the 2SC-s state for which $\Delta_r=\Delta_g=\Delta_b$ and 
$\mu_8=0$. This figure strongly suggests that the 2SC-s state is, 
in fact, the absolute minimum of the mean field thermodynamical 
potential. 
  
Another important feature of the 2SC-s state concerns the corresponding 
12 quasiparticle modes. It turns out that for this state the dispersion 
relations and degeneracy factors are given by 
\begin{eqnarray} 
E_0^\pm & = & |\vec p\,| \pm \mu_B/3 \qquad\qquad\qquad\;\;\, 
[\,\times 2\,]\ ,  \\ 
E_\Delta^\pm & = & \sqrt{(|\vec p\,| \pm \mu_B/3)^2 + \Delta^2} 
\qquad [\,\times 4\,]\ , 
\end{eqnarray} 
where $\Delta = \sqrt{3} \Delta_s$. This means that the system 
contains two gapless modes, just like in the case of the conventional 
2SC-$b$ state. However, in 
the present case the gapless modes cannot be identified with the 
original $u$ and $d$ quarks of ``blue'' (unpaired) color in the 
original $\{r,g,b\}$ color basis but arise as a combination of all 
three color states. 
 
In summary, we have shown in this letter that in the case of color 
neutrality the ground state of 2SC quark matter should be 
constructed in a ``democratic'' way, so that color symmetry is not 
broken by the choice of the orientation of the condensate vector 
in color space. For this state, the condition of color neutrality 
is fulfilled in a trivial way, since the penalty induced by 
otherwise necessary color chemical potentials is avoided. We have 
shown that this 2SC-s ground state is stable against fluctuations, 
thus the problem observed in \cite{hjz} is solved. The 2SC-s state 
can serve as a starting point for considering hadronic 
correlations on the superconducting QCD vacuum, where due to the 
entanglement of the quark color states in the new basis color 
neutral quark and diquark excitations arise along the radial 
direction besides colored excitations in the tangential plane. The 
conclusions drawn so far should not be qualitatively altered when 
the additional constraint of electrical neutrality is imposed, 
which is important for applications in compact stars. Flavor 
asymmetry induced by this constraint could possibly inhibit the 
formation of the 2SC state in regions of the neutron star matter 
phase diagram. However, provided that a phase transition to 2SC 
quark matter is accomplished, it should be described by the 2SC-s 
state introduced in this paper. This issue, together with other 
extensions of the present work, will be explicitly discussed in 
forthcoming publications. 
 
{\bf Acknowledgements. } D.B.~acknowledges discussions with 
Michael Buballa and the inspiring Color Superconductivity seminar 
of the Virtual Institute held at Frankfurt and Darmstadt. This 
work has been supported in part by CONICET and ANPCyT (Argentina), 
under grants PIP 02368, PICT00-03-08580 and PICT02-03-10718, and 
by a scientist exchange program between Germany and Argentina 
funded jointly by DAAD under grant No.~DE/04/27956 and ANTORCHAS 
under grant No.~4248-6.

\end{document}